\documentclass[proof]{WileyASNA-v1}

\articletype{Article Type}%

\received{15 October 2021}
\revised{15 October 2021}
\accepted{15 October 2021}

  \newcommand\gsim{\lower.6ex\hbox{$\buildrel>\over\sim$}}
\newcommand\lsim{\lower.6ex\hbox{$\buildrel<\over\sim$}}

\begin{document}

\title{What defines a compact symmetric object? A carefully vetted sample of compact symmetric objects}

\author[1]{Anthony C. S. Readhead*}

\author[2,3]{Sebastian Kiehlmann}

\author[4]{Matthew L. Lister}

\author[1]{Sandra O'Neill}

\author[1]{Timothy~J. Pearson}

\author[5]{Evan Sheldahl}

\author[6]{Aneta Siemiginowska}

\author[5]{Gregory B. Taylor}

\author[7]{Peter N. Wilkinson}

\authormark{READHEAD \textsc{et al}}

\address[1]{\orgdiv{Owens Valley Radio Observatory}, \orgname{California Institute of Technology}, \orgaddress{\state{Pasadena, California 91125}, \country{USA}}}

\address[2]{\orgdiv{Institute of Astrophysics}, \orgname{Foundation for Research and Technology-Hellas}, \orgaddress{\state{GR-71110 Heraklion}, \country{Greece}}}

\address[3]{\orgdiv{Department of Physics}, \orgname{Univ. of Crete}, \orgaddress{\state{Heraklion}, \country{Greece}}}

\address[4]{\orgdiv{Department of Physics and Astronomy}, \orgname{Purdue University}, \orgaddress{\state{525 Northwestern Avenue, West Lafayette, IN 47907}, \country{USA}}}

\address[5]{\orgdiv{Department of Physics and Astronomy}, \orgname{University of New Mexico}, \orgaddress{\state{Albuquerque, NM 87131}, \country{USA}}}

\address[6]{\orgdiv{Harvard-Smithsonian Center for Astrophysics}, \orgaddress{\state{60 Garden Street, Cambridge, MA 02138}, \country{USA}}}

\address[7]{\orgdiv{Jodrell Bank Centre for Astrophysics}, \orgname{University of Manchester}, \orgaddress{\state{Oxford Road, Manchester M13 9PL}, \country{UK}}}

\corres{A.C.S Readhead, California Institute of Technology, Pasadena, California 91125, USA. \email{acr@caltech.edu}}

\abstract{Compact Symmetric Objects  (CSOs), young jetted-AGN of overall projected size <1 kpc, are of great interest due to their youth and evolution. The classification was introduced to distinguish between $\sim$95\% of powerful compact extragalactic radio sources in flux density limited samples that are dominated by asymmetric emission due to relativistic beaming from jets aligned close to the line of sight, and $\sim$5\% of objects that are  not. The original classification criteria were: (i) overall projected diameter smaller than ~1 kpc, (ii) identified center of activity, and (iii) symmetric jet structure about the center. There is confusion and erosion of the value of the CSO classification due to misclassifications.  Many jets contain compact bright features outside core, resulting in a GPS total spectrum and a ``compact double'' appearance, and some objects with jet axes aligned close to the line of sight appear symmetric because the approaching jet is projected on both sides of the core.  To eliminate the confusion, we propose adding  (iv) slow radio variability and  (v) low apparent velocity of bright features moving along the jets to the above CSO criteria. We are compiling a catalog of CSOs using these five criteria to eliminate the confusion of Doppler boosting.}

\keywords{active galaxies, compact symmetric objects, Jetted-AGN}

%%\fundingInfo{Funding info text.}

\maketitle

\footnotetext{\textbf{Abbreviations:} AGN, active galactic nuclei; CSO, compact symmetric object; IPS, interplanetary scintillation; CSS, compact steep spectrum source; GPS, Gigahertz peaked spectrum source; MPS, Megahertz peaked spectrum source; PS, peaked spectrum source; HFP, high frequency peakers}

\begin{figure*}[ht!]
\centering
\includegraphics[width=0.8\textwidth]{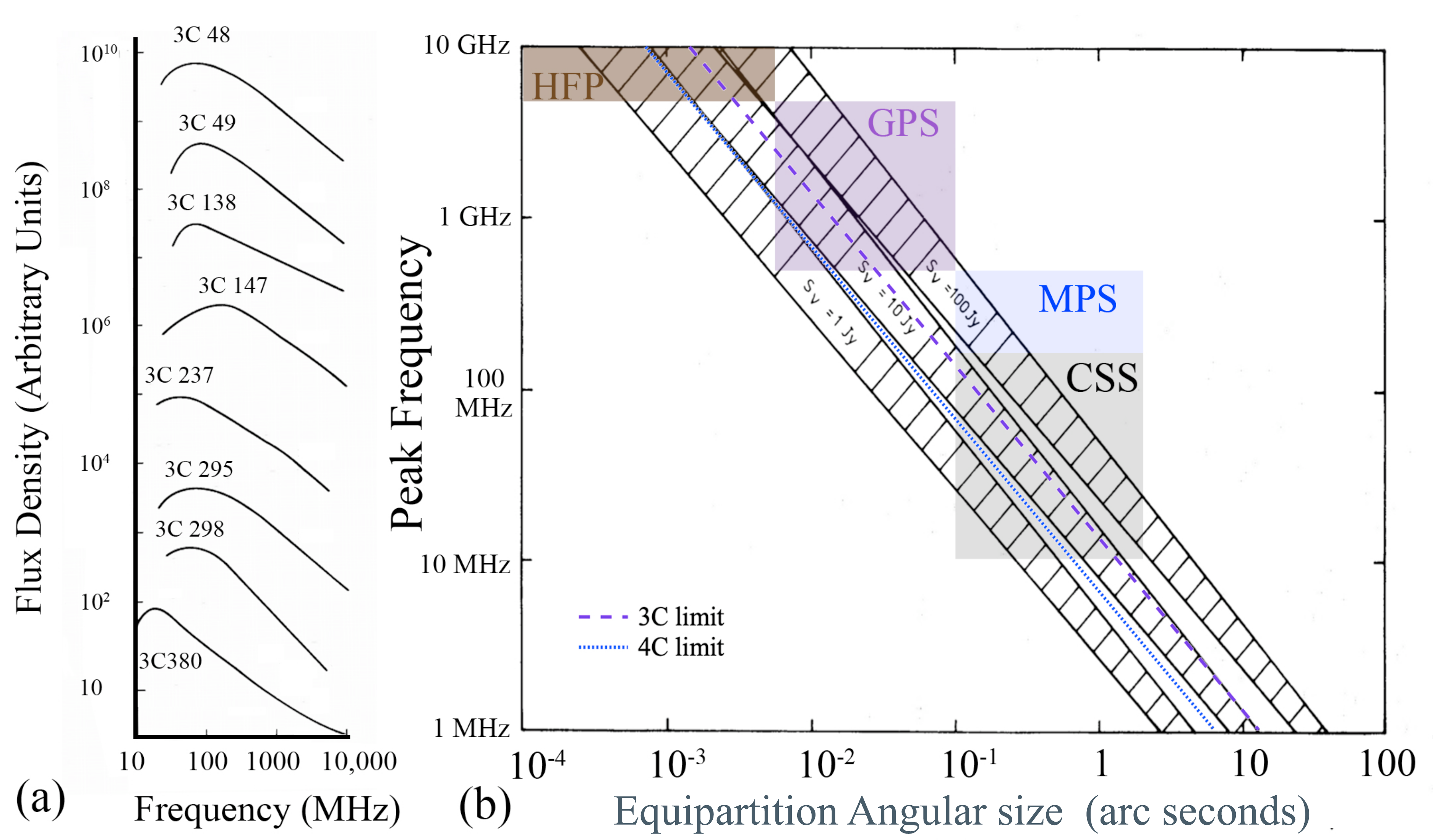}
\caption{Adapted from \citet{1977MNRAS.180..539S}: (a) The CSS objects that show strong interplanetary scintillation have spectra exhibiting peaks due to synchrotron self-absorption between $\sim$30 MHz and $\sim$100 MHz; (b) the peak frequency, for a source exhibiting synchrotron self-absorption, as a function of equipartition angular size. The hatched regions indicate the peak frequencies  for sources with peak flux densities 1 Jy, 10 Jy, and 100 Jy. The upper and lower bounds of the hatched regions in each case refer to a source with $\alpha = -1.5$  and $\alpha = -0.3$, respectively, for an object at $z$ = 0.7. The purple dashed and blue dotted lines represent an object at $z$ = 0.7, with $\alpha=-0.75$, at the 178 MHz flux density limit  of the 3C and 4C catalogs, respectively. The brown, purple, blue and grey shaded regions indicate the typical locations of HFP, GPS, MPS and CSS objects, all of which peak at frequencies above 10 MHz and therefore belong to the PS class of AGN (see text).}
\label{plt:spectra}
\end{figure*}

\section{Introduction}\label{intro}

The goal of looking for Compact Symmetric Objects (CSOs) is to find small, young double-jetted active galactic nuclei (AGN). CSOs have to have small angular sizes ($\ll 1$ arc second), but any sample selected by angular size alone is polluted by relativistically beamed sources that have small angular size due to projection. Of course there will be physically small sources that are beamed towards us that would be CSOs if seen at a different angle. But they are difficult or impossible to distinguish from the larger beamed sources, so we choose to exclude all sources showing indications of beaming. The Compact Symmetric Object (CSO) class of  compact radio sources was defined by \citet{1994ApJ...432L..87W} to address this problem by discriminating between ``core-jet'' objects, in which the morphology is determined by relativistic beaming emission, and CSOs in which it is not. Unfortunately, as explained in this paper, a number of compact radio sources in which the emission \textit{is} strongly beamed towards the observer have been classified as CSOs or CSO candidates.  This has undermined the rationale for the CSO classification and led to much confusion regarding CSOs.  In this paper we review the background to the CSO classification, with the discovery of compact doubles \citep{1980ApJ...236...89P} and of recurrent activity \citep{1990A&A...232...19B} in compact doubles, justify the need for additional criteria for classifying CSOs, and  describe a large study undertaken to compile a carefully vetted catalog of CSOs.

\section{The PS CLASSIFICATION OF AGN}\label{sec:nomenclature}
The review of compact extragalactic radio  sources by \citet{2021A&ARv..29....3O} shows that they are key to studies of relativistic jets, star formation, AGN feedback, and structure formation, and hence embrace a range of subjects of fundamental importance to astrophysics and cosmology. Compact extragalactic radio sources have been of great interest since their discovery through long baseline interferometry by \citet{1962MNRAS.124..477A}, who showed that 3C 48, 3C 119, 3C 147 and 3C 237 all have angular sizes $< 3$ arc seconds.  This was shortly followed by the discovery of interplanetary scintillation by \citet{1964Natur.203.1214H}, who showed that 3C 48, 3C 119, 3C 138 and 3C 237 had angular sizes $<$ 1 arc second. \citet{1976MNRAS.176..571R} 
showed that, in a complete sample of 181 3CR sources, 66 had scintillation indices greater than 0.4, indicating that a large fraction of their flux density arises in a region smaller than 1 arc second. \citet{1977MNRAS.180..539S} showed  that the brightest strongly scintillating steep spectrum sources exhibit peaked spectra between $\sim$50 MHz and $\sim$100 MHz, as seen in Fig. 1a\footnote{
We define the spectral index, $\alpha$, by $S \propto \nu^\alpha$} and that the emission regions in these objects are close to equipartition between the particle and magnetic field energy densities. \citet{2021A&ARv..29....3O} have suggested grouping high-frequency peaked (HFP) objects  \citep{2017ApJ...836..174C, 2000A&A...363..887D} with gigahertz peaked spectrum (GPS) objects and objects exhibiting peaks in the 100 MHz -- 1 GHz range (MPS) into a single class of peaked spectrum (PS) objects. This is a very useful new classification, given the continuity in the observed peak frequency -- angular size correlation from MHz to GHz frequencies.  

The equipartition angular size \citep{1977MNRAS.180..539S,2021ApJ...907...61R}, $\psi_{_{\rm eq}}$ in arc seconds, in a source showing a synchrotron self-absorption peak in the spectrum at $S$ Jy and $\nu$ MHz is
\begin{equation}\label{psi1}
\psi_{_{\rm eq}} = 1.67 \times  r^{-{1 \over17}} S^{8 \over 17}  \nu^{-{{35+ 2\alpha} \over 34}}   (1+z)^{{15-2\alpha} \over 34}
 F(\alpha) \, ,
\end{equation}
 where $\psi$ is the source angular diameter at the peak frequency of a uniform brightness disk, $r$ is the comoving coordinate distance to the source in gigaparsecs,  and $F(\alpha)$ is given in \citet{1977MNRAS.180..539S}.
 This shows that, as can be seen in Fig. 1b, any CSS source with peak flux density greater than $\sim$ 1 Jy and angular size at the peak less than $\sim$ 1 arc second will have a spectral turnover above $\sim$ 10 MHz.

According to \citet{1977MNRAS.180..539S}, all bright ($S>$1 Jy) compact steep spectrum (CSS) radio sources must also exhibit peaks in their spectra above $\sim$10 MHz, and we therefore suggest enlarging the definition of PS objects proposed by \citet{2021A&ARv..29....3O} to include CSS objects as well, even if no peak has been observed in the spectrum down to the lowest frequency observed, because there is no known physical difference between these two classes of object, apart from size, which is in both cases significantly less than the size of the host galaxy.

\section{Compact Doubles, Recurrent Activity, and Compact Triples}\label{sec:cds}
The discovery of compact double sources \citep{1980ApJ...236...89P,1982A&A...106...21P} came as a complete surprise since up to that time \textit{all} compact sources resolved by VLBI had been one-sided ``core-jet'' objects in which the emission is strongly affected by relativistic beaming. As shown by \citet{1978Natur.276..768R,1980IAUS...92..165R}, and recognized by \citet{1984RvMP...56..255B}, this provided a unifying model of steep- and flat-spectrum radio sources as well as of radio galaxies and quasars.

\citet{1980ApJ...236...89P} attributed the compact double morphology to youth and suggested that these might be the precursors of the much larger double sources that dominate low-frequency surveys. The prediction that these are young radio sources has been spectacularly confirmed, as can be seen by the kinematic ages of the compact objects listed in Table 3 of \citet{2021A&ARv..29....3O}.

In their VLBI survey, \citet{1988ApJ...328..114P} discriminated between compact doubles with steep \textit{vs.} flat high frequency spectra, and drew attention to the fact that those with steep high frequency spectra show little variability or polarization and are primarily galaxies.

The next surprise relating to compact doubles came with the discovery by \citet{1990A&A...232...19B} of evidence for an earlier phase of activity in the compact double source 0108+388.  As discussed in \citet{2021A&ARv..29....3O}, there is now much evidence for recurrent activity amongst compact radio sources.

In  the meantime the story of compact double sources had become complicated with the discovery of the first compact triple source \citep{1984IAUS..110..131R}, which was followed by the discovery of more compact triples \citep{1992ApJ...396...62C,1994ApJ...425..568C}, which raised the question of how best to classify these objects.

\section{Why Do we Need the CSO classification?}\label{sec:why}

There are clearly serious problems of interpretation when relying solely on the GPS, CSS, compact double, and compact triple  classifications because of the intermingling of relativistically beamed ``core-jet'' objects having axes aligned close to the line of sight and objects with axes closer to the plane of the sky.  For this reason \citet{1994ApJ...432L..87W} defined the CSO class  based on the following criteria: (i) the projected size must be less than 1 kpc,  (ii) the structure must be symmetric, as revealed by twin jets, twin hot spots and/or twin lobes, (iii) the center of activity
 must either be detected directly or inferred from convincing images of jets and/or lobes and/or hotspots straddling the nucleus or its inferred position.  
 \citet{1994ApJ...432L..87W} point out a further distinguishing characteristic of CSOs: (iv) they show low variability, but this attribute is rarely invoked in classifying CSOs. We wish to draw attention to its importance, and in addition to advance a fifth criterion for CSO disqualification: (v) CSOs should not exhibit superluminal motion in excess of 2.5$c$ for the reasons discussed below.

\begin{figure*}
    \centering
    \includegraphics[width=0.95\linewidth]{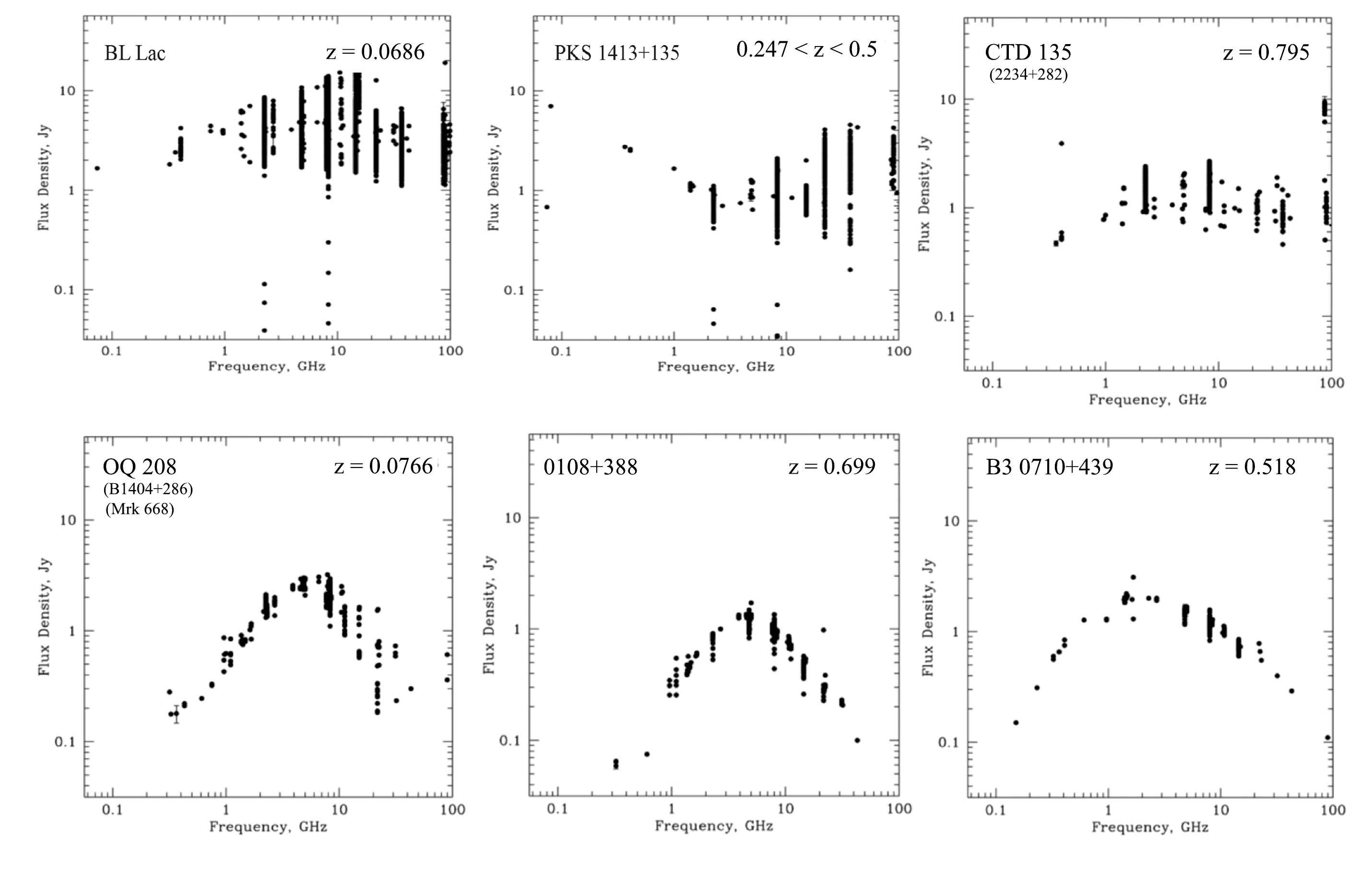}
    \caption{Radio spectra taken from the MOJAVE website \citep{2019ApJ...874...43L}. All of the data shown here are total flux densities, so the spread of values at each frequency are an indication of the level of variation of the flux density at that frequency over the last three decades. The first object is the blazar BL Lac, the archetype of its class. The other five objects have all been classified as CSOs or CSO candidates, and include the archetypal CSO  B3 0710+439.     Apart from the typical CSO (B3 0710+439), which shows very little variability at all frequencies and on all timescales, the other five objects show strong variability above 3 GHz on timescales of weeks to years. The similarity between PKS 1413+135 and BL Lac, which is a typical blazar, argues strongly for the jet in PKS 1413+135 being aligned closely with the line of sight, and therefore not a CSO.}
    \label{spectra1}
\end{figure*}

\subsection{The Radio Spectra and Variability of Blazars {\it vs.} CSOs}\label{spectra}
Significant fractional flux density variations in quasars on timescales of months to years \citep{1965Sci...148.1458D,1965Sci...150...63M,1965Natur.208..275A} are what originally  led \citet{1966Natur.211..468R} to propose that there are emission regions in the nuclei of quasars that are moving at relativistic speeds towards the observer.  These objects were also known to have flat radio spectra.  This also led Spiegel \citep{1980ARA&A..18..321A} to propose the name blazar for this class of jetted-AGN, in which the jets are aligned close to the line of sight.  

Since the CSO classification was introduced specifically by \citet{1994ApJ...432L..87W} to distinguish the small subset of $\sim 5\%$ of the radio sources in high frequency surveys with compact nuclear structure having jets aligned close to the plane of the sky from the majority of $\sim 95 \%$ of the sources with jets aligned close to the line of sight \citep{1988ApJ...328..114P}, the topics of variability and radio spectrum are a good starting point for our discussion of the criteria for CSO classification.

Multi-epoch radio spectra of  BL Lac and five objects classified as CSOs  or CSO candidates (PKS 1413+135, CTD 135, OQ 208, 0108+388 and B3 0710+439) are shown in Fig. \ref{spectra1}. These are all taken from the MOJAVE website\footnote{https://www.physics.purdue.edu/MOJAVE/}. All the points plotted are total flux densities, so the spread of points at a given frequency is an indicator of variability in total flux density in these objects.  It is immediately clear that the radio spectra and variability of PKS 1413+135 and CTD 135 above 3 GHz are extremely similar to the those of BL Lac, in which the jet axis is closely aligned with the line of sight,  and very dissimilar to the spectrum and variability of the archetypal CSO B3 0710+439, in which the jet axis is closer to the plane of the sky than to the line of sight.  

In order to address the problems raised above regarding the CSO classification we need to introduce a criterion based on variability  to the original criteria set out by \citet{1994ApJ...432L..87W}. Clearly, variability can be used to disqualify an object from the CSO class if it exhibits flux density variations on a timescale that is significantly shorter than the light travel time across the emission region.

A prime example is provided by the detailed and comprehensive study of the jetted-AGN OQ 208 by \citet{2013A&A...550A.113W}. The spectrum and variability of OQ 208 are shown in Fig. \ref{spectra1} where it can be seen that the high frequency spectrum is steep and the variability at 15 GHz  is about a factor 3.  \citet{2013A&A...550A.113W} show that the 14.5 GHz flux density, as measured at the University of Michigan Radio Observatory, varied from $\sim 2$ Jy, in 1982, to $\sim$ 0.7 Jy in 2010.  So this flux density variation of a factor 3 occurred over a period of 28 years. The variations are dominated by the component ``NE1'' identified by \citet{2013A&A...550A.113W} which is unresolved and has size set by the primary beam of $\sim 1.5$ light years $\times$ $0.7$ light years.  So relativistic beaming is not required to explain  the variability.  Thus we classify OQ 208 as a bona fide CSO.

We considered using a spectral index cutoff of $\alpha = -0.6$ above 3 GHz to distinguish between blazars aligned close to the line of sight and CSOs aligned close to the plane of the sky, but decided against this as it would exclude CSOs such as TXS 0128+554 in which the viewing angle relative to the jet axis is $\sim 50^\circ$, and the flux density of the core on the approaching side is  mildly Doppler boosted by $\sim$1.2 \citep{2020ApJ...899..141L}. In such objects the Doppler boosting does not dominate and hence the emitted flux densities are close to those observed.

An example of an object that has been suggested as a CSO candidate, which should definitely be rejected on  the grounds of its variability  is CTD 135, which is also shown in Fig. \ref{spectra1}.  This object has been discussed in detail by \citet{2016AN....337...65A}.

This brings us to the case of PKS 1413+135, which is what started us on this study. PKS 1413+135  is one of the most peculiar jetted-AGN known \citep{2021ApJ...907...61R}, but its strange properties are outside the scope of this paper. It has frequently been discussed in the literature as a CSO and sometimes found to be unlike the other members of this class -- see, for example \citet{2010ApJ...713.1393W}. From the above discussion of the variability and spectrum of CSOs and from these features of PKS 1413+135 as exhibited in Fig. \ref{spectra1}, it should be clear that we reject this object from the CSO class on account of its rapid variability.

\subsection{Apparent Velocities of Components in CSO Jets}\label{velocities}
 
Apparent velocities, $v_{\rm app}$, of components  in CSO jets relative to the core, apart from the exemplar misclassified object PKS 1413+135,  are always consistent with the velocity of light, i.e. $v_{\rm app}\lesssim c$ \citep{1996ApJ...460..612R, 2016AJ....152...12L, 2016MNRAS.459..820T}. PKS 1413+135 stands out in this regard since it exhibits some components in its jet moving at speeds modestly in excess of the speed of light. The maximum apparent speed in PKS 1413+135 is $v_{\rm app} = (1.72  \pm 0.011) c$ \citep{2019ApJ...874...43L}, for an assumed redshift of $z$ = 0.247, and a factor 1.58 higher for an assumed redshift of 0.5 \citep{2021ApJ...907...61R}. For comparison the maximum apparent speeds of components in the jets of two typical CSOs measured by MOJAVE are 0108+388: $v_{\rm app,mas} = (0.83 \pm 0.15) c$; 0710+439: $v_{\rm app,max} = (1.03 \pm 0.32) c$; and for 1946+708 it is $v_{\rm app,max} = (1.088 \pm 0.011) c$ \citep{2009ApJ...698.1282T}.  

We adopt a limit of 2.5$c$ for the speed of any jet components relative to the core in CSOs because larger values would imply a jet viewing angle of less than 45$^\circ$ and Lorentz factor $>$ 3.

\section{A Revised Classification for CSOs}\label{sec:revised}

The above considerations led us to undertake an intensive review of the literature.  We began by searching for papers that mentioned CSOs in their titles and/or abstracts, and as we progressed we added all papers discussing CSOs that we found. We also searched the GPS and CSS literature for objects that were promising CSO candidates based on criteria other than their spectra.
In this way we reviewed 143 papers and considered 2077 objects classified as CSOs or CSO candidates and/or GPS or CSS objects.  Each of the 143 papers we reviewed were read by two of us and any objects  of interest were added to a data base that included all relevant radio images, spectra, and lightcurves. Objects that were not obviously CSOs or CSO candidates, or  that were not clear rejects, were referred to the group for regular discussions. 
We classified the objects into four distinct groups: (i) bona fide CSOs, (ii) ``Class A candidates'' for which the currently available data does not provide enough evidence for a classification as a bona fide CSO, but that are good candidates for follow-up VLBI  observations; (iii) ``Class C candidates'' are similar to ``Class A'' but are less promising candidates for VLBI follow-up, (iv) objects that are definitely or almost definitely not CSOs.

As would be expected, the clear rejects were predominantly ``core-jet'' objects. In this process the difficult cases were discussed by all of us.  Most of the rejects were due to low dynamic range images based on VLBI snapshot observations with very sparse (u,v) coverage. The result of this exercise was that 81 objects were classified as bona fide CSOs, 167 objects were classified as ``Class A'' CSO candidates, 913 objects were classified as ``Class C'' CSO candidates, and 916 objects were rejected from the CSO candidate list. 

The most striking result of this exercise was that 62 of the rejected objects had previously been classified as CSOs or CSO candidates. This illustrates how confused the literature  has become since 1994, and highlights the need for a clearer set of criteria for membership of the CSO class of AGN if this classification is to be of real value.

A further aspect of our study was recurrent activity in CSOs, which has been of great interest since the seminal paper of \citet{1990A&A...232...19B}. To study this aspect we also carried out a parallel study of compact sources for which there is evidence of recurrent activity.  Here we studied 63 papers and considered recurrent activity in 206 objects.  The result was that we found 9 bona fide CSOs that show recurrent activity, and 6 bona fide  medium symmetric objects (MSOs), i.e. objects with largest size in the range 1 kpc -- 10 kpc that show recurrent activity.

These results are now being written up in two papers, one of which will provide a carefully vetted catalog of bona fide CSOs. This is not a complete catalog in any sense, our object has been to identify as many bona fide CSOs as possible, so we have been conservative in our selection.  This is to ensure that relativistic beaming is not a factor when studying CSOs. The second paper discusses recurrent activity in CSOs.

We have been granted time on the VLBA for observations   to make high-quality dual-frequency images of all the ``Class A'' CSO candidates, and we pursuing  high-quality dual-frequency VLA maps of the CSO six objects that show recurrent activity. This may well point to the need for deep VLA images of all bona fide CSOs in a search for fainter evidence of recurrent activity in these  objects in which a recent burst of activity is clearly evident.

\section*{Acknowledgments}
S.K.\ acknowledges support from the European Research Council (ERC) under the European Unions Horizon 2020 research and innovation programme under grant agreement No.~771282.
S.O.\ gratefully acknowledges the support of the Caltech Summer Undergraduate Research Fellowship program.  A.S.\ was supported by NASA contract NAS8-03060 to the Chandra X-ray Center.

\bibliography{Main}

\section*{Author Biography}

\begin{biography}
{}%{\includegraphics[width=60pt,height=70pt,draft]{empty}}
{\textbf{Anthony Readhead}  is the Robinson Professor of Astronomy (Emeritus) at the California Institute of Technology.}
\end{biography}

\end{document}